\documentclass[aps, prl, twocolumn, letterpaper, superscriptaddress]{revtex4-2}

\usepackage{amsmath}
\usepackage{amssymb}
\usepackage{xspace}
\usepackage{color}
\usepackage{graphicx}
\usepackage{hyperref}
\usepackage{ulem}
\usepackage{ifpdf}
\ifpdf
\fi

\newcommand{\eq}[1]{(\ref{#1})}
\newcommand{\Eq}[1]{Eq.~(\ref{#1})}

\newcommand{\Fig}[1]{Fig.~\ref{#1}}

\newcommand{\eg}{{e.g.,\/}\xspace}
\newcommand{\ie}{{i.e.,\/}\xspace}
\newcommand{\etal}{{\it et~al.\/}\xspace}

\newcommand{\pd}{\partial}
\newcommand{\mc}[1]{\mathcal{#1}}
\newcommand{\bd}[1]{\boldsymbol{#1}}
\newcommand{\msection}[1]{{\it #1. -- }}

\newcommand{\avg}[1]{\left\langle{#1}\right\rangle}
\newcommand{\exb}{{E\times B}}
\newcommand{\ExB}{{\bd{E}\times \bd{B}}}

\begin{document}

\title{Intrinsic toroidal rotation driven by turbulent and neoclassical processes in tokamak plasmas from global gyrokinetic simulations}

\author{Hongxuan Zhu}
\affiliation{Princeton Plasma Physics Laboratory, Princeton, NJ 08540}
\affiliation{Department of Astrophysical Sciences, Princeton University, Princeton, NJ 08544}

\author{T. Stoltzfus-Dueck}
\affiliation{Princeton Plasma Physics Laboratory, Princeton, NJ 08540}

\author{R. Hager}
\affiliation{Princeton Plasma Physics Laboratory, Princeton, NJ 08540}

\author{S. Ku}
\affiliation{Princeton Plasma Physics Laboratory, Princeton, NJ 08540}

\author{C. S. Chang}
\affiliation{Princeton Plasma Physics Laboratory, Princeton, NJ 08540}
\begin{abstract}
Gyrokinetic tokamak plasmas can exhibit intrinsic toroidal rotation driven by the residual stress. While most studies have attributed the residual stress to the parallel-momentum flux from the turbulent $\boldsymbol{E}\times\boldsymbol{B}$ motion, the parallel-momentum flux from the drift-orbit motion (denoted $\Pi_\parallel^D$) and the $\ExB$-momentum flux from the $\ExB$ motion (denoted $\Pi_{\exb}$) are often neglected. Here, we use the global total-$f$  gyrokinetic code XGC to study the residual stress in the core and the edge of a DIII-D H-mode plasma. Numerical results show that both $\Pi_\parallel^D$ and $\Pi_{\exb}$ make up a significant portion of the residual stress. In particular, $\Pi_\parallel^D$ in the core is higher than the collisional neoclassical level in the presence of turbulence, while in the edge it represents an outflux of counter-current momentum even without turbulence. Using a recently developed ``orbit-flux'' formulation, we show that the higher-than-neoclassical-level $\Pi_\parallel^D$ in the core is driven by turbulence, while the outflux of counter-current momentum from the edge is mainly due to collisional ion orbit loss. These results suggest that $\Pi_\parallel^D$ and $\Pi_\exb$ can be important for the study of intrinsic toroidal rotation.
\end{abstract}

\maketitle
Tokamak plasmas can rotate toroidally without external momentum input, which is important for future reactors where internal fusion heating is not expected to generate net momentum. Such intrinsic toroidal rotation is driven by the residual stress, which is a momentum flux independent from the toroidal-rotation velocity and its gradient. The gyrokinetic approach is often used to find the residual stress in turbulent plasmas, but its determination can be difficult because turbulence will transport equal amounts of co- and counter-current momentum, so the net momentum flux is zero unless there is an asymmetry in the parallel direction. Therefore, studies of the residual stress have been active for many years \cite{Peeters11,Diamond13,Parra15,SD19,Camenen09,McDevitt09,Waltz11,Ku12,Buchholz14,Lee14,Grierson17,SD17,Hornsby17,Hornsby18,Ball18,Scott10,Brizard11,Garbet13}.

In an electrostatic gyrokinetic plasma, radial transport comes from the drift-orbit motion $\bd{v}_D$ and the turbulent $\ExB$ motion $\bd{v}_E$. While most studies have attributed the residual stress to the parallel-momentum flux from $\bd{v}_E$ (the ``fluid stress''), the parallel-momentum flux from $\bd{v}_D$ (the ``kinetic stress'') and the $\ExB$-momentum flux from $\bd{v}_E$ (the ``toroidal Reynolds stress'') are often neglected in the core. In particular, the kinetic stress is usually assumed to be at a small collisional neoclassical level. However, as will be discussed in this paper,  part of the kinetic stress can be driven by turbulence, which is already observed by several global gyrokinetic simulations \cite{Abiteboul11,Sarazin11,Idomura14,Idomura17,Garbet17} and studied from a qualitative theory \cite{Garbet17}. Numerically, the residual stress is often studied in the local geometry, where the volume-integrated kinetic stress and Reynolds stress vanish due to radial periodicity, but this radial boundary condition no longer exists in the  global geometry. The kinetic stress has also been emphasized for the edge rotation \cite{Chang08,Seo14,Muller11,Muller11b}, which is affected by not only turbulence, but also complicated factors such as the realistic geometry with a magnetic X point \cite{Chang08,Seo14,Muller11,Muller11b,SD12,SD15,Fedorczak12,Peret22}, interactions with neutrals \cite{Helander03,Stroth11}, and ion orbit loss \cite{Itoh88,Shaing89,Chankin93,Miyamoto96,Chang02,Ku04,deGrassie09,deGrassie12,Brzozowski19}.  With the advancing computing power, global gyrokinetic simulations with realistic geometry could provide new physics insights for this topic.

In this paper, we use the global total-$f$ particle-in-cell gyrokinetic code XGC \cite{XGC} to study the residual stress in a DIII-D H-mode plasma. Both the core and the edge are studied through whole-volume plasma simulation from the magnetic axis to the wall. We initiate the plasma  with zero rotation velocity and study  the self-generated momentum fluxes. The gyrocenter toroidal angular momentum (TAM) density consists of the parallel-flow part $\mc{L}_\parallel$ and the $\ExB$-flow part $\mc{L}_\exb$. Their corresponding radial TAM fluxes are denoted by $\Pi_\parallel$ and $\Pi_\exb$, respectively. Numerical results show that both $\Pi_\parallel^D$  (the component of $\Pi_\parallel$ from $\bd{v}_D$) and  $\Pi_\exb$ make up a significant portion of the residual stress. Using a recently developed ``orbit-flux'' formulation \cite{SD20,SD21,Zhu22,Zhu23}, we quantitatively show how $\Pi_\parallel^D$ is driven not only by collisions, but also by turbulence in the core, as well as by collisional ion orbit loss in the edge. Similar results are also found in the core of a larger machine ITER, as discussed toward the end. These results suggest that $\Pi_\parallel^D$ and $\Pi_\exb$ can be important for the study of intrinsic toroidal rotation.

\begin{figure}
    \centering
    \includegraphics[width=1\columnwidth]{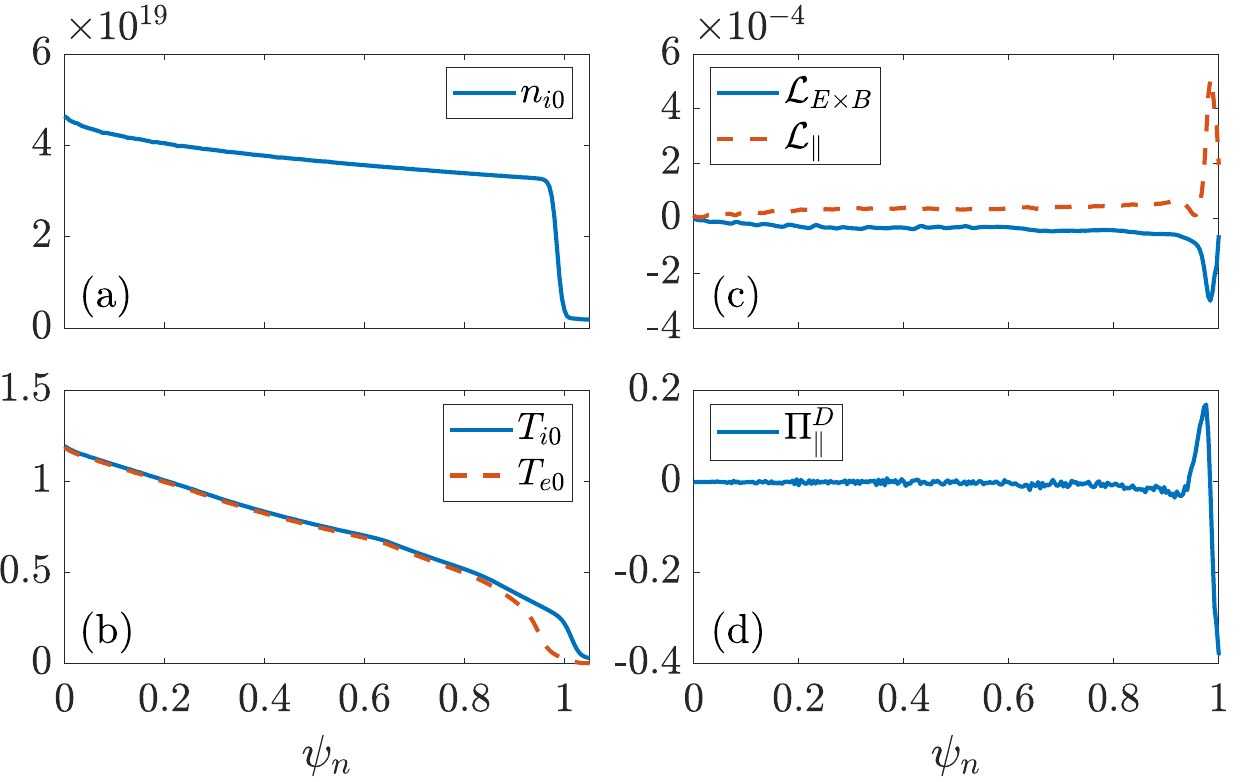}
    \caption{(a) and (b): the equilibrium density (in units ${\rm m}^{-3}$) and temperature (in units keV) as a function of normalized poloidal flux $\psi_n$, and $n_{e0}=n_{i0}$ due to quasineutrality. (c) and (d): the TAM density (in units $\mathrm{kg/(m\cdot s)}$) and flux (in units $\mathrm{N\cdot m}$) at $t=0.4$ms from the neoclassical XGCa simulation.}
    \label{fig:D3D_equil}
\end{figure}

\msection{Simulation setup} We simulate deuterium gyrokinetic ions and drift-kinetic electrons. Their equilibrium density and temperature are adapted from DIII-D shot number 141451 \cite{Seo14,Muller11,Muller11b} and are functions of the poloidal magnetic flux $\psi$ (\Fig{fig:D3D_equil}). The gyrocenter coordinates are cylindrical position $\bd{R}=(R,\varphi,z)$, magnetic moment $\mu$, and parallel momentum $p_\parallel$. The Hamiltonian for species $s$ is $H=p_\parallel^2/2m_s+\mu B+Z_se\hat{J}_0{\Phi}$, where $\hat{J}_0\Phi$ is the gyroaveraged electrostatic potential. Define $v_\parallel=p_\parallel/m_s$, $\hat{\bd{b}}=\bd{B}/B$, $\bd{B}^*=\bd{B}+(m_sv_\parallel/Z_se)\nabla\times\hat{\bd{b}}$, and $B_\parallel^*=\hat{\bd{b}}\cdot\bd{B}^*$, then the gyrocenter trajectories are given by
\begin{equation*}
B_\parallel^*\dot{\bd{R}}=v_\parallel\bd{B}^*+(Z_se)^{-1}\hat{\bd{b}}\times \nabla H,\quad B_\parallel^*\dot{p}_\parallel=-{\bd{B}}^*\cdot\nabla H.     
\end{equation*}
Separating $H$ into an axisymmetric part $\bar{H}=p_\parallel^2/2m_s+\mu B+Z_se\hat{J}_0\bar{\Phi}$ and a nonaxisymmetric part $\tilde{H}=Z_se\hat{J}_0\tilde{\Phi}$, we have $\dot{\bd{R}}=v_\parallel\hat{\bd{b}}+\bd{v}_D+\bd{v}_E$, where $v_\parallel\hat{\bd{b}}+\bd{v}_D$ is the parallel and the drift-orbit motion and $\bd{v}_E$ is the $\ExB$ motion from $\tilde{\Phi}$. Note that $\bd{v}_D$ includes not only the grad-$B$ and the curvature drift, but also the $\ExB$ drift from $\bar{\Phi}$.

The ``total-$f$'' numerical scheme is used for the whole-volume plasma simulation \cite{Ku16}, where ``$f$'' refers to the gyrocenter distribution $F_s$, which evolves according to 
\begin{equation}
\label{eq:XGC_vlasov}
d_tF_s=\pd_t F_s+\dot{\bd{R}}\cdot\nabla F_s+\dot{p}_\parallel \pd_{p_\parallel}F_s=C_s+S_s+N_s,
\end{equation}
and is allowed to significantly deviate from the equilibrium Maxwellian distribution. Here, $C_s$ describes collisions \cite{Yoon14,Hager16}, $S_s$ describes heating, and $N_s$ describes neutral ionization and charge exchange \cite{Ku18}. In our simulations, a 1MW heating is applied to ions in the core to sustain turbulence, and neutral dynamics are included in the edge and the scrape-off layer. 

Neither $S_s$ nor $N_s$ generate net momentum in our simulations. Then, the gyrokinetic equation \eq{eq:XGC_vlasov} has a local gyrocenter TAM conservation relation \cite{Scott10,Brizard11,SD17,Garbet13,Abiteboul11}
\begin{equation}
\label{eq:TAM_conservation}
\pd_t(\mc{L}_\parallel+\mc{L}_\exb)=-\pd_V(\Pi_\parallel+\Pi_\exb),
\end{equation}
where $V(\psi)$ is the volume inside the flux surface $\psi$. The TAM densities are calculated as  
$\mc{L}_\parallel=\sum_s\langle\int d^3 v F_sp_{s\varphi}\rangle$ and $\mc{L}_\exb=-(dV/d\psi)^{-1}\int dt\sum_sZ_se\Gamma_s$, where $\avg{\dots}$ is the flux-surface average, $p_{s\varphi}=-m_sv_\parallel\hat{\bd{b}}\cdot R^2\nabla\varphi$ is the TAM from parallel motion, and $\Gamma_s=\langle \int d^3vF_s\dot{\bd{R}}\cdot\nabla V\rangle$ is the radial gyrocenter flux. The sign of $p_{s\varphi}$ is chosen so that a positive (negative) TAM density corresponds to a co- (counter-) current toroidal rotation. The TAM fluxes are calculated as $\Pi_{\parallel}=\sum_s\langle\int d^3v F_sp_{s\varphi}(\dot{\bd{R}}\cdot\nabla V)\rangle$ and $\Pi_{\exb}=-\int dV\sum_s\langle\int d^3v F_s\pd_\varphi H\rangle$. Since radial transport comes from both $\bd{v}_D$ and $\bd{v}_E$, we write 
\begin{equation}
\Gamma_s=\Gamma_s^D+\Gamma_s^E,\quad \Pi_\parallel=\Pi_\parallel^D+\Pi_\parallel^E
\end{equation}
to emphasize their separate contributions.

\begin{figure}
    \centering    \includegraphics[width=0.8\columnwidth]{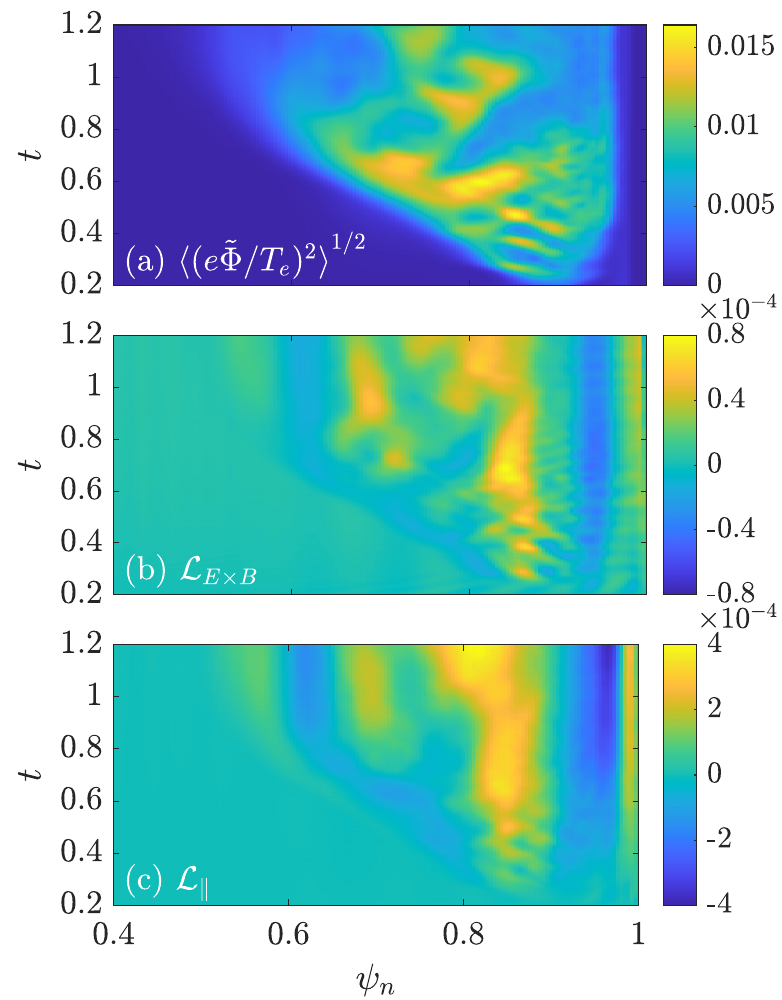}
    \caption{XGC1 simulation results showing (a) amplitude of the turbulent-fluctuations, (b) $\mc{L}_\exb$, and (c) $\mc{L}_\parallel$ as a function of $t$ (in units ms) and $\psi_n$. The XGCa solutions of $\mc{L}_\parallel$ and $\mc{L}_\exb$ are subtracted to remove their large peaks in the edge.}
    \label{fig:D3D_turb}
\end{figure}

\msection{Simulation results} First, we use the axisymmetric version of XGC (XGCa) to simulate a neoclassical plasma, where $\Pi_\parallel^E=\Pi_\exb=0$ but $\Pi_\parallel^D$ can be nonzero due to collisions. Starting from a local Maxwellian $F_s$, the plasma relaxes to a quasisteady state at $t=0.4$ms, when the TAM densities and flux are shown in Figs.~\ref{fig:D3D_equil}(c) and \ref{fig:D3D_equil}(d). In the core, $\mc{L}_\exb<0$  due to a negative neoclassical radial electric field $E_r$, while $\mc{L}_\parallel>0$  from the parallel return flow that balances the poloidal $\ExB$ and diamagnetic flow. The neoclassical-level $\Pi_\parallel^D$ is very small in the core, so the TAM density is conserved at each flux surface, $\mc{L}_\parallel\approx -\mc{L}_\exb$. In the edge, $\mc{L}_\exb$ has a counter-current peak at $\psi_n\approx 0.99$ due to the H-mode edge $E_r$ well. Correspondingly,  $\mc{L}_\parallel$ has a co-current peak, but the relation $\mc{L}_\parallel\approx -\mc{L}_\exb$ is no longer satisfied due to a dipolar $\Pi_\parallel^D$ in the edge. Throughout the simulation, the edge $\mc{L}_\parallel$ shifts in the counter-current direction at the pedestal top ($\psi_n<0.98$) and in the co-current direction toward the last closed flux surface ($\psi_n=1$) according to \Eq{eq:TAM_conservation}. 

Next, we use the 3D version of XGC (XGC1) to simulate a turbulent plasma and the results are shown in \Fig{fig:D3D_turb}. Turbulence is active in the core but decays in the edge due to the $H$-mode $E_r$ well. In the core, turbulence-driven $\mc{L}_\exb$ and $\mc{L}_\parallel$ have similar radially wavelike structures. Note that here $\mc{L}_\parallel$ and $\mc{L}_\exb$ have the same sign, which is different from the XGCa solution $\mc{L}_\parallel\approx -\mc{L}_\exb$. In the edge, turbulent intensity is weak so that the TAM flux is dominated by $\Pi_\parallel^D$, and the corresponding edge rotation is also similar to the XGCa solution. The observed edge $\Pi_\parallel^D\approx -0.3\mathrm{N\cdot m}$ is comparable to that inferred from experiments \cite{Muller11,Muller11b}, and our simulation results in the edge are qualitatively consistent with the results using a previous version of XGC with a different setup \cite{Seo14}. 

The above results showed that both neoclassical and turbulent processes can generate residual TAM fluxes and toroidal rotation in our simulations. In the following, we study the physics behind these momentum fluxes.

\begin{figure}
    \centering    \includegraphics[width=1\columnwidth]{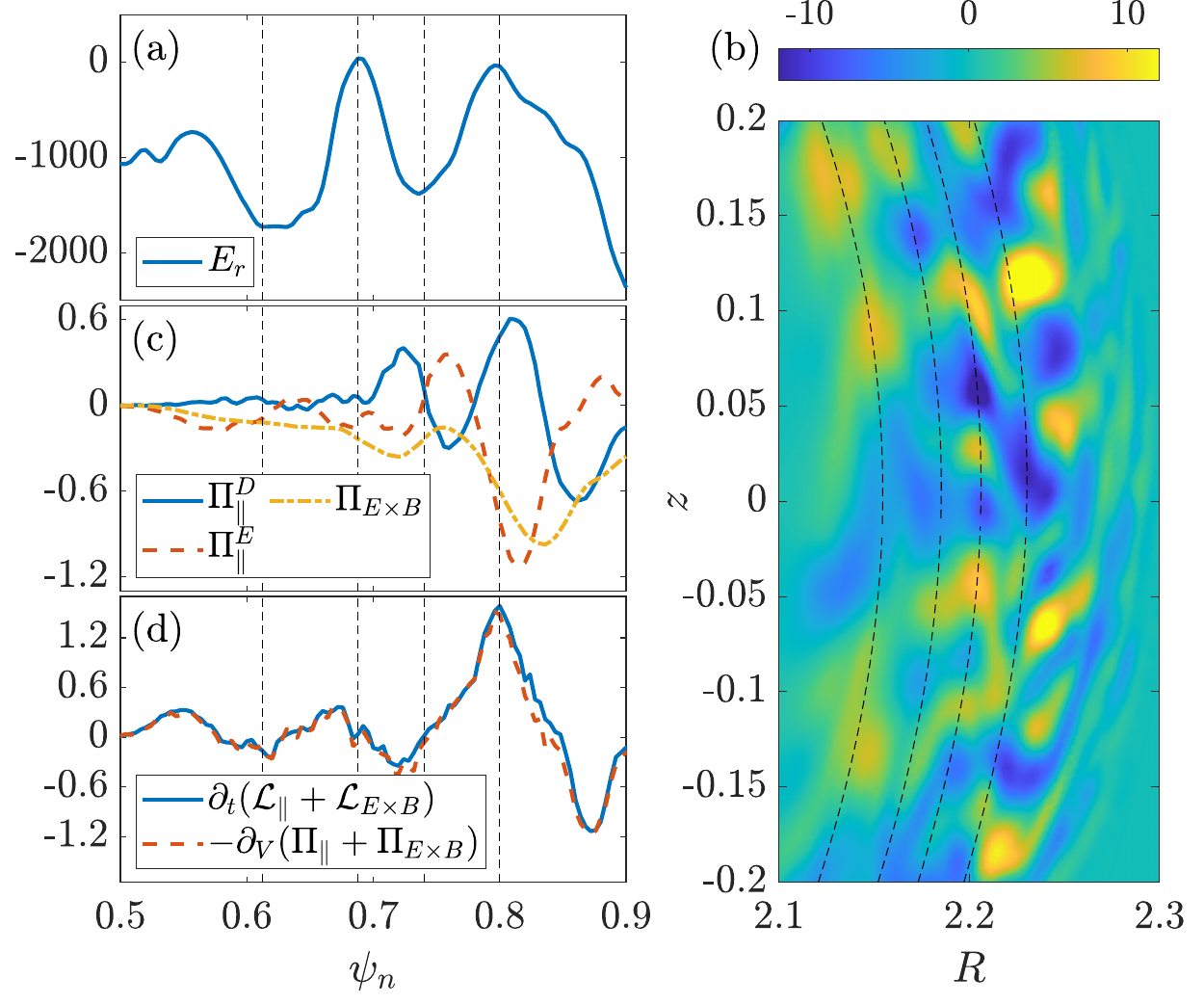}
    \caption{(a) $E_r$ (in units V/m) at $t=1$ms along the outboard midplane. The black dashed lines are flux surfaces where $\pd_r E_r\approx 0$. (b) $\tilde{\Phi}$ (in units V) near the outboard midplane. (c) The corresponding TAM fluxes. (d) Comparison of the TAM conservation relation \eq{eq:TAM_conservation} with numerical results.}
    \label{fig:D3D_mflux}
\end{figure}

\msection{Core momentum fluxes} Figure~\ref{fig:D3D_mflux}(a) shows $E_r$, which varies radially and drives differential poloidal rotation known as zonal flows. Since $\mc{L}_\exb$ is proportional to $E_r$, the observed correlation between $\mc{L}_\exb$ and $\mc{L}_\parallel$ can be understood as the correlation between zonal flows and toroidal rotation, which was also seen in other global gyrokinetic simulations \cite{Wang09,Wang10,Wang11,Wang17}. As shown in Figs.~\ref{fig:D3D_mflux}(b) and (c), turbulent eddies are tilted according to the local zonal-flow shear, and the corresponding $\Pi_\parallel^E$ and $\Pi_\exb$ oscillate radially. Meanwhile, $\Pi_\parallel^D$ is larger than the neoclassical solution in \Fig{fig:D3D_equil}(d) and tends to be out of phase with $\Pi_\parallel^E$, in agreement with other global gyrokinetic-simulation results \cite{Abiteboul11,Sarazin11,Idomura14,Idomura17,Garbet17}. Therefore, all the three TAM fluxes should be considered in order to correctly predict the toroidal-rotation evolution  in the core [\Fig{fig:D3D_mflux}(d)].

We found these TAM fluxes significant in the sense that $|v_t\Pi/aQ_i|$ can be as large as 0.5, meaning they can drive toroidal rotation up to a nonnegligible fraction of the ion thermal velocity $v_{t}$ \cite{Hornsby17,Hornsby18}. (Here, $a$ is the minor radius and $Q_i$ is the ion heat flux). It is well known that the zonal-flow shear can produce finite correlation between poloidal and parallel wave spectra and hence a nonzero $\Pi_\parallel^E$ \cite{Dominguez93,Gurcan07}. However, studies often assumed that $\Pi_\exb$ is smaller than $\Pi_\parallel^E$ by a factor $k_r\rho_iB_\theta/B$ and $\Pi_\parallel^D$ is at a small collisional neoclassical level \cite{Peeters11}. Our results showed that these assumptions are not always valid, and we focus on the origin of $\Pi_\exb$ and $\Pi_\parallel^D$ in the following. 

\begin{figure}
    \centering \includegraphics[width=1\columnwidth]{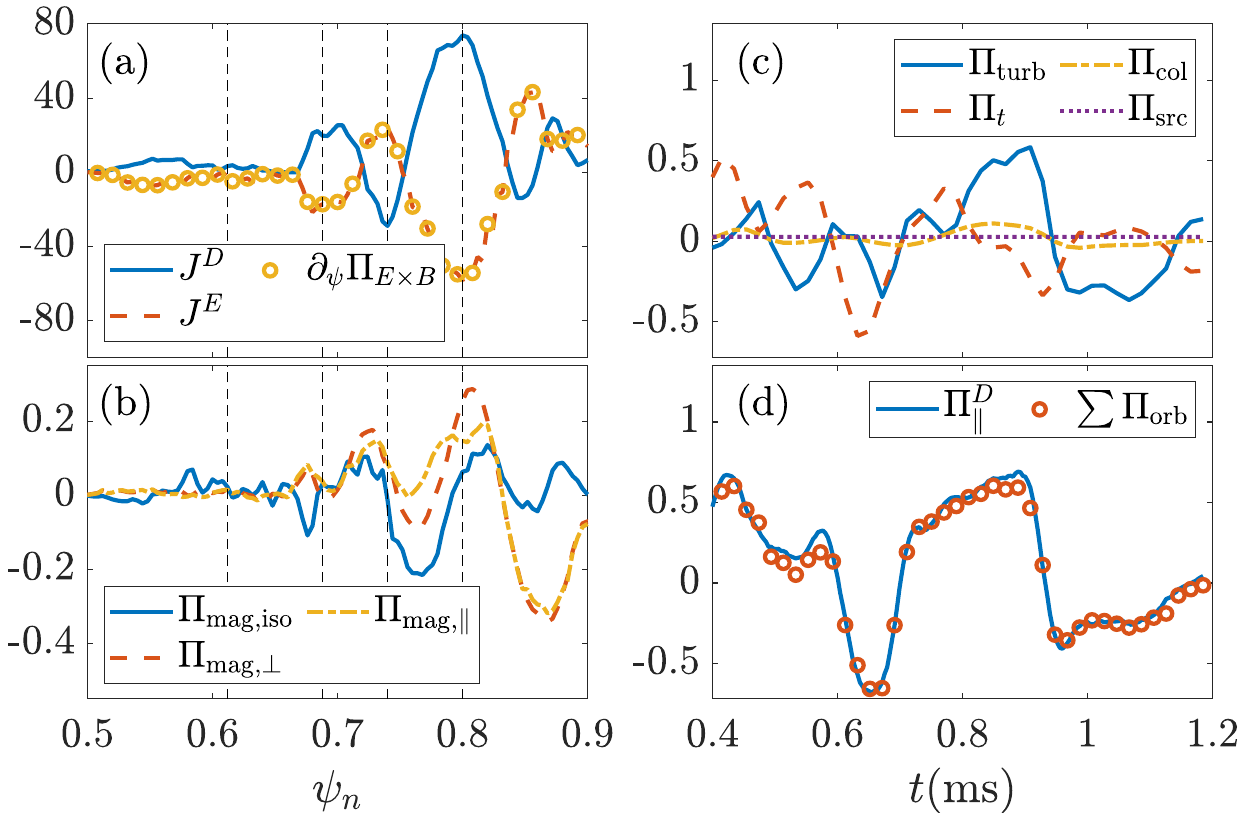} 
    \caption{(a) The  radial currents and  $\pd_\psi\Pi_\exb$ (in units A) at $t=1$ms. The black dashed lines are the flux surfaces plotted in \Fig{fig:D3D_mflux}. (b) The isothermal and non-isothermal parts \eq{eq:PiD_iso} of $\Pi_\parallel^D$ in \Fig{fig:D3D_mflux} . (c) Orbit-flux calculations \eq{eq:orbit_PiD} for momentum  fluxes across the $\psi_n=0.76$ core flux surface. (d) Comparison with the direct calculation of $\Pi_\parallel^D$ from XGC1 using \Eq{eq:flux_PiD_orig}.}
    \label{fig:D3D_flow}
\end{figure}

Using the relation $\pd_\varphi=Rb_\varphi\hat{\bd{b}}\cdot\nabla+B^{-1}\hat{\bd{b}}\times\nabla\psi\cdot\nabla$ and assuming $k_\parallel\ll k_\perp$ for turbulence, one can show that
\begin{equation}
\label{eq:approx_Piexb}    \pd_V\Pi_\exb\approx(dV/d\psi)^{-1}J^E,
\end{equation}
where $J^E=\sum_sZ_se\Gamma_s^E$ is the turbulent radial current. This approximation is numerically verified in \Fig{fig:D3D_flow}(a), and can be interpreted as the proportionality between toroidal and poloidal projection of the Reynolds stress. Note that the drift-orbit current $J^D=\sum_sZ_se\Gamma^D_s$ balances $J^E$ so that the total gyrocenter current is small. By comparing Figs.~\ref{fig:D3D_mflux}(a) and \ref{fig:D3D_flow}(a), these radial currents also oscillate with the zonal flow. Since a positive gyrocenter current drives $E_r$ in the negative direction and vice versa,  the zonal flow is driven by $J^E$ and damped by $J^D$. As $E_r$ forms according to $J^E$, toroidal rotation driven by $\Pi_\exb$ will have the same radial profile as $E_r$ according to \Eq{eq:approx_Piexb}. This is a novel explanation for the correlation between toroidal rotation and zonal flows.

Although $\bd{v}_D$ contains the $\ExB$ drift from $\bar{\Phi}$, the corresponding TAM flux is small in our simulations, so most of $\Pi_\parallel^D$ is from the magnetic (grad-$B$ and curvature) drift  of ions, $\bd{v}_{\rm mag}\approx(\mu B+m_iv_\parallel^2)\hat{\bd{b}}\times\nabla\ln B/Z_ie$. Write $\mu B+m_iv_\parallel^2=2T_{i0}+(\mu B-T_{i0})+(m_iv_\parallel^2-T_{i0})$ as the contributions from isothermal processes and deviation to Maxwellian distribution in the perpendicular and parallel directions, the kinetic stress $\Pi_\parallel^D\approx\langle\int d^3v F_ip_{i\varphi}\bd{v}_{\rm mag}\cdot\nabla V\rangle$ can then be written as
\begin{equation}
   \Pi_\parallel^D=\Pi_{{\rm mag, iso}}+\Pi_{{\rm mag},\perp}+\Pi_{{\rm mag},\parallel},\label{eq:PiD_iso}
\end{equation}
with the three terms from $2T_{i0}$,  $\mu B-T_{i0}$, and $m_iv_\parallel^2-T_{i0}$, respectively. It is straightforward to show that $\Pi_{\rm mag,iso}\propto-T_{i0}\langle\gamma_\parallel\sin\theta\rangle$, where $\gamma_\parallel=\int d^3vF_iv_\parallel$ is the ion parallel flux density and $\theta$ is the poloidal angle. Therefore, a nonzero $\Pi_{\rm mag,iso}$ arises due to the up-down asymmetry in $\gamma_\parallel$. As discussed in Ref.~\cite{Garbet17}, such asymmetric $\gamma_\parallel$ can be driven by the divergence of the turbulent radial flux, \ie $\Pi_{\rm mag,iso}\propto\pd_r\Gamma_i^E$, which explains the radially oscillatory behavior of $\Pi_\parallel^D$ in \Fig{fig:D3D_mflux}. However, as shown in \Fig{fig:D3D_flow}(b), both $\Pi_{{\rm mag},\perp}$ and $\Pi_{{\rm mag},\parallel}$ are comparable to $\Pi_{\rm mag,iso}$, so the qualitative theory from Ref.~\cite{Garbet17} alone cannot explain the turbulence-driven $\Pi_\parallel^D$ in our simulations. Further, we found the contributions to $\Pi_{{\rm mag},\perp}$ and $\Pi_{{\rm mag},\parallel}$ from temperature fluctuations to be small, so they must come from higher-order moments in the ion distribution.

\msection{Turbulent origin of $\Pi_\parallel^D$ in the core} Although the observed $\Pi_\parallel^D$ cannot be simply explained from the low-order fluid moments of $F_i$, we can still numerically illustrate the turbulent origin of $\Pi_\parallel^D$  using a recently developed ``orbit-flux'' formulation \cite{SD20,SD21}. By definition, the kinetic stress at flux surface $\psi$ is
\begin{equation}
\label{eq:flux_PiD_orig}
    \Pi_\parallel^D=\frac{2\pi}{m_i^2}\oint \sqrt{g}\,d\theta\, d\varphi\int dp_\parallel  d\mu B_\parallel^* F_i p_{i\varphi}\bd{v}_D\cdot\nabla\psi,
\end{equation}
where $\sqrt{g}=|\nabla\psi\times\nabla\theta\cdot\nabla\varphi|^{-1}$. Since drift-orbit motion $v_\parallel\hat{\bd{b}}+\bd{v}_D$ conserves the canonical TAM $\mc{P}_\varphi=p_{i\varphi}-Z_ie\psi$ and the energy $\bar{H}$, we can use $(\mu,\mc{P}_\varphi,\bar{H})$ to label all drift orbits that cross the flux surface  $\psi$. Changing variables from $(p_\parallel,\theta)$ to $({\mc{P}_\varphi,\bar{H}})$, we obtain
\begin{equation}
\label{eq:flux_PiD}
    \Pi_\parallel^D=\frac{2\pi}{Z_iem_i^2}\int d\mu\,d\mc{P}_\varphi\,d\bar{H}\oint d\varphi(F_i^{\rm out}-F_i^{\rm in})p_{i\varphi}.
\end{equation}
Here, it is assumed that each drift orbit  crosses the flux surface twice, one at the incoming point and the other at the outgoing point, and we define $F_i^{\rm in}$ and $F_i^{\rm out}$ to be the ion distribution at these two points, respectively. For each drift orbit, $\Delta F_i=F_i^{\rm out}-F_i^{\rm in}$ can be calculated as an orbit integration from the incoming point to the outgoing point at fixed time $t$:
\begin{equation}
\label{eq:DeltaF}
    \Delta F_i=\int d\tau(C_i+S_i+N_i-{\tilde{\dot{\bd{R}}}\cdot\nabla F_i}-{\tilde{\dot{p}}_\parallel\pd_{p_\parallel}F_i}-\pd_tF_i),
\end{equation}
where $\tilde{\dot{\bd{R}}}=\bd{v}_E$, $\tilde{\dot{p}}_\parallel=-\bd{B}^*\cdot\nabla\tilde{H}/B_\parallel^*$, and the integration is along drift orbits parameterized by $\tau$. Combining \eq{eq:flux_PiD} and \eq{eq:DeltaF}, we write $\Pi_\parallel^D$ as the summation of ``orbit fluxes'':
\begin{equation}
\label{eq:orbit_PiD}
    \Pi_\parallel^D=\Pi_{\rm col}+\Pi_{\rm src}+\Pi_{\rm neut}+\Pi_{\rm turb}+\Pi_{t}.
\end{equation}
A similar procedure can be applied to $\Gamma_i^D$ to obtain
\begin{equation}
\label{eq:orbit_GammaD}
    \Gamma_i^D=\Gamma_{\rm col}+\Gamma_{\rm src}+\Gamma_{\rm neut}+\Gamma_{\rm turb}+\Gamma_{t}.
\end{equation}
Equations \eq{eq:orbit_PiD} and \eq{eq:orbit_GammaD} are called ``orbit-flux'' formulations, which show that $\Pi_\parallel^D$ and $\Gamma_i^D$ are not only driven by collisions (which is the focus of the conventional neoclassical theory), but also by heating, neutral dynamics, turbulence, and time evolution of the plasma along collisionless drift orbits. Our simulated plasma is in the low-collisionality banana regime, with mostly collisionless drift orbits up to the last closed flux surface.

We numerically implemented this formulation in XGC \cite{Zhu22,Zhu23}. As an example, we look at $\Pi_\parallel^D$ at the $\psi_n=0.76$ flux surface and the results are in \Fig{fig:D3D_flow}(c) and (d). The dominant contribution to  $\Pi_\parallel^D$ are the turbulent term $\Pi_{\rm turb}$ and the associated time evolution of the plasma  $\Pi_t$, while effects from collisions and heating are small.  Also, the orbit-flux calculation \eq{eq:orbit_PiD} agrees well with the direct calculation of $\Pi_\parallel^D$ in XGC1 using \eq{eq:flux_PiD_orig}, which demonstrated that it is implemented with good numerical accuracy. These results quantitatively confirmed that the higher-than-neoclassical-level $\Pi_\parallel^D$ in the XGC1 simulation is indeed driven by turbulence. 

\begin{figure}
    \centering    \includegraphics[width=1\columnwidth]{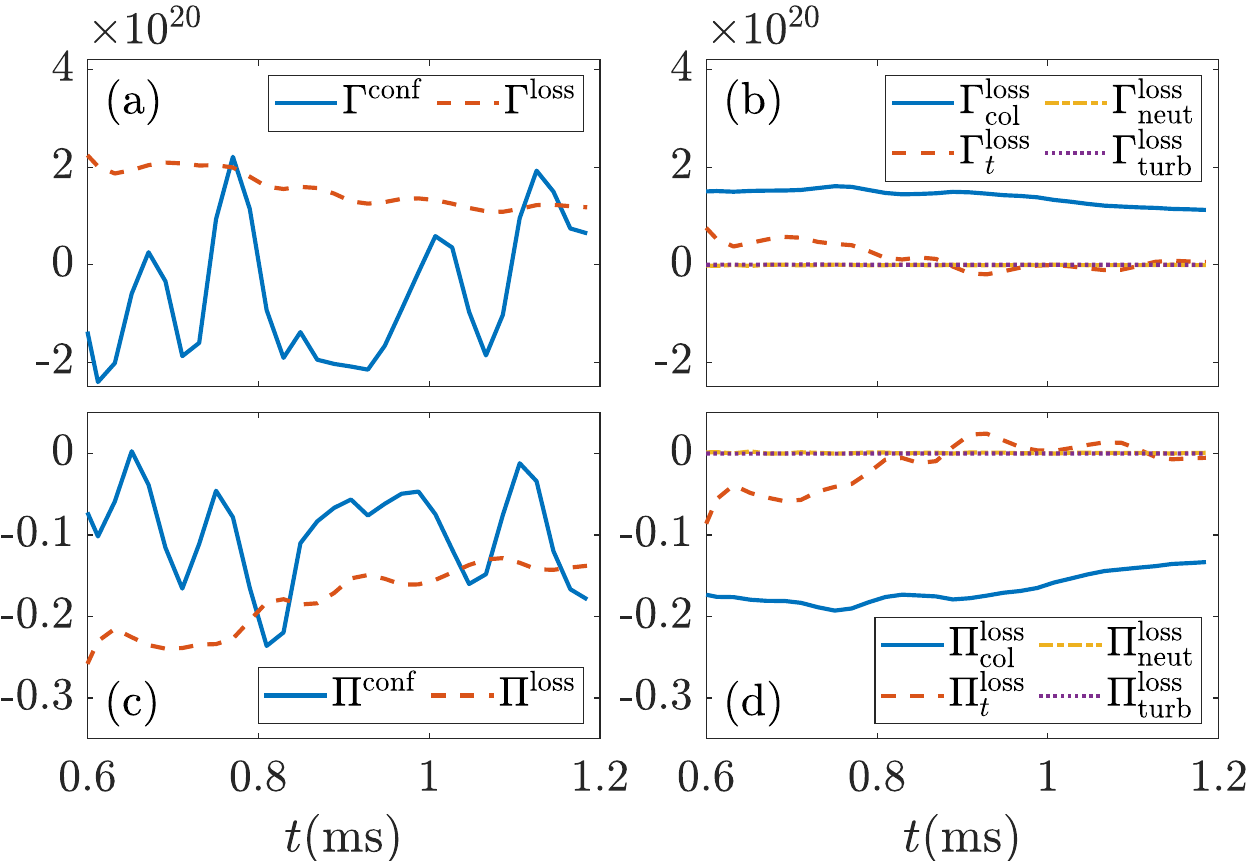}
    \caption{(a) and (c): The particle and momentum fluxes across the $\psi_n=0.992$ edge flux surface. Shown are the separate contribution from confined orbits and loss orbits. (b) and (d): orbit-flux calculations \eq{eq:orbit_PiD} and \eq{eq:orbit_loss} for the loss orbits.}
    \label{fig:D3D_edge_flux}
\end{figure}

\msection{Edge momentum fluxes} In the edge region of our simulation, turbulence is weak and $\Pi_\parallel^D$ is mainly driven by neoclassical processes. Note that the edge is  subject to ion orbit loss, where some drift orbits do not form closed loops but connect the confined region to the divertor leg or the vessel wall. Considering the separate contributions from loss orbits and the remaining confined orbits (which form closed loops), we write orbit fluxes as, \eg
\begin{equation}
\label{eq:orbit_loss}
    \Gamma_{\rm col}=\Gamma_{\rm col}^{\rm loss}+\Gamma_{\rm col}^{\rm conf}, \quad \Pi_{\rm col}=\Pi_{\rm col}^{\rm loss}+\Pi_{\rm col}^{\rm conf}.
\end{equation} 
Results for an edge flux surface $\psi_n=0.992$ are shown in \Fig{fig:D3D_edge_flux}. For the particle flux, we find $\Gamma^{\rm loss}>0$ and $\Gamma^{\rm conf}<0$. In other words, while gyrocenter ions leave the plasma following the loss orbits, they also enter the plasma following the confined orbits. For the momentum flux, however, both $\Pi^{\rm loss}$ and $\Pi^{\rm conf}$ are counter-current and they add up to  $\Pi_\parallel^D\approx -0.3\mathrm{N\cdot m}$ in the edge. These results are consistent with each other, namely, most loss orbits are counter-current and the remaining confined orbits are overall co-current, so that both $\Gamma^{\rm loss}>0$ and $\Gamma^{\rm conf}<0$ result in counter-current momentum fluxes. The loss-orbit fluxes are mainly caused by collisional scattering of ions into the loss orbits, while effects from turbulence and neutrals are small. Therefore, the outgoing counter-current momentum fluxes in the edge is mainly from collisional ion orbit loss within our simulation. Finally, note that the self-consistent orbit-loss driven $\Pi_\parallel^D$ determines $\pd_t\mc{L}_\parallel$ in the edge, which is different from simple orbit-loss models that determine $\mc{L}_\parallel$ itself \cite{deGrassie09,deGrassie12}.

\msection{Conclusions} In summary, global total-$f$ gyrokinetic simulations showed that $\Pi_\parallel^D$ is higher than the collisional neoclassical level in the presence of turbulence, and both $\Pi_\parallel^D$ and $\Pi_\exb$ make up a significant portion of the residual stress in a DIII-D H-mode plasma. Using the orbit-flux formulation, we identified the mechanisms that drive $\Pi_\parallel^D$, including turbulence in the core and collisional ion orbit loss in the edge. It is often assumed that $\bd{v}_D$ gives rise only to neoclassical transport, which is driven solely by collisions and is smaller than the turbulent transport from $\bd{v}_E$. Our results showed that this assumption is not always valid, because part of the radial transport from $\bd{v}_D$ can be driven by turbulence. In Supplemental Material \cite{supplemental}, we provide an ordering estimate for $\Pi_\parallel^D$ and argue that it can be comparable to $\Pi_\parallel^E$; we also report similar results in simulations of electrostatic turbulence in a larger machine ITER. These results suggest that $\Pi_\parallel^D$ and $\Pi_\exb$ can be important for the study of intrinsic toroidal rotation, and global gyrokinetic simulations could lead to further new physics insights for this topic.

\begin{acknowledgments}
This work was supported by the U.S. Department of Energy under Contract Number DE-AC02-09CH11466. The United States Government retains a non-exclusive, paid-up, irrevocable, world-wide license to publish or reproduce the published form of this manuscript, or allow others to do so, for United States Government purposes. Funding to R. Hager, S. Ku and C. S. Chang is provided via the SciDAC-4 program. The simulations presented in this article were performed on computational resources managed and supported by Princeton Research Computing, a consortium of groups including the Princeton Institute for Computational Science and Engineering (PICSciE) and the Office of Information Technology’s High Performance Computing Center and Visualization Laboratory at Princeton University. This research used resources of the National Energy Research Scientific Computing Center, which is supported by the Office of Science of the U.S. Department of Energy under Contract No. DE-AC02-05CH11231. Digital data can also be found in DataSpace of Princeton University \cite{Data}.
\end{acknowledgments}

\end{document}